Reply to cond-mat/0601101
V.N.Zavaritsky
Loughborough University, Loughborough, United Kingdom
Kapitza Physics Institute & General Physics Institute, Moscow, Russia

*The recent PRB 72,094503(2005)., henceforth referred as Ref.[1], experimentally resolves the intrinsic shape of the c-axis current-voltage characteristics (IVC) of HTSC and demonstrates that at sufficiently high heat loads the heating-induced IVC nonlinearities exceed the intrinsic ones so radically that the latter might be safely ignored, as is evident from the fact that Newton's law of cooling and Ohm's law describe the 'intrinsic tunnelling' spectra quantitatively*

*The author of the comment ignores the experimental findings by Ref.[1] and promotes a brush-like IVC, which is claimed to be free of self-heating. I will show that this claim lacks substantiation as the IVC is definitely not free from heating and* that the self-heating cause of IVC-2 is indirectly admitted by the author of the comment. *I will further show that the data selected for this comment in fact provide additional experimental evidence in favour of the major conclusions by Ref.[1] in particular of the extrinsic cause of the key findings by intrinsic tunnelling spectroscopy.*

Heat W, dissipated in a sample, escapes through its surface area (A) and causes significant heating if the heat load P=W/A exceeds the critical value $P_c$, which depends on the experimental environment. Notably, $P_c$ is close to 1W/cm^2 for liquid helium and is significantly smaller for helium vapour at a comparable temperature. Heating is probably the most common problem in low temperature research and a particularly harsh limiting factor for the study of current-voltage characteristics (IVC). Self-heating of superconductors is particularly well studied experimentally and theoretically (see Ref.[2] for a comprehensive review). In particular, heating often causes IVC nonlinearities and transforms a single-valued IVC into a multi-valued characteristic with regularly spaced branches.

The findings summarised by the authors of Ref.[2] are particularly relevant to high temperature superconductors (HTSC) because the exceptionally poor thermal and electrical conductivities of HTSC makes them particularly prone to local heating. However, unlike other studies of HTSC, the heating issues in 'intrinsic tunnelling' devoted to the brush-like IVC were misinterpreted or ignored until recently. Particularly confusing claims arise from 'intrinsic tunnelling spectroscopy' (IJT), which postulates that HTSCs factually represent natural stacks of atomic-scale intrinsic superconductor-insulator-superconductor (SIS) Josephson junctions. IJT further postulates the intrinsic cause of the IVC features built by the heat loads in excess of kilowatts per cm^2. Such loads, however, exceed the corresponding $P_c$ by 4-6 orders of magnitude (Refs.[1,3]), which indicates that unlike conventional spectroscopy, the heating in IJT is not a small perturbation but a principal cause of IVC nonlinearity (see Ref.[4] for details). Indeed, the systematic studies summarised in Ref.[1] show *that at sufficiently high heat loads the heating-induced IVC nonlinearities exceed the plausible intrinsic ones so radically that the latter might be safely ignored. The experimental IVC in such circumstances is primarily determined by the normal state resistance, $R_N(T)$, while the mean temperature, T, of the self heated sample is appropriately described by Newton's Law of Cooling (1701),*

$$T=T_B+P/h, \qquad (Eq.1)$$

where $T_B$ is the temperature of the coolant medium (liquid or gas) and *h* is the heat transfer coefficient, which depends neither on A nor T, see Refs.[1,5,7] for details. The extrinsic cause of the key IJT findings was established by Refs.[1,5,7-9] with this parameter-free description. The consistency of this description was reaffirmed by independent measurements by Ref.[6].

Ref.[1] presents experimental tools for distinguishing intrinsic features from extrinsic ones. In particular, Ref.[1] addresses a generic IVC hidden by heating artefacts and shows that it is Ohmic above $T_c$ while the brush-like part is reasonably described by:

$$V\_\# = R\_\#(I - I^{\wedge}*); \qquad (Eq.2)$$

Here the differential resistance of a resistive branch (R_#) is proportional to its number, #, and represents a fraction of the normal state resistance R_N of the same sample measured under conditions of complete suppression of its superconductivity. As is seen from Eq.2, the generally non-Ohmic response is close to Ohmic if the offset current I^* is sufficiently small. This common case is illustrated in Fig.1a which shows typical IJT IVC (which were originally claimed as "Evidence for Coexistence of the Superconducting Gap and the Pseudogap" by the authors of Ref.[10]) re-plotted as a sample resistance, R=V/I, normalised by its value at P=0, versus the heat load, P=IV/A. As is seen from Fig.1(a), there is a well defined threshold level, P_c, below which R(P) is flat, while it drops rapidly at P>P_c. According to Refs.[1,4.7,8,9], the R(P) curves in the latter case are caused by Joule self-heating and hence must obey Eq.(1). Indeed, as is seen from Fig.1(b), the parameter-free Eq.(1) collapses all IVC into a single curve which reproduces the measured R(T) and allows an estimate of the heat transfer coefficient h=32Wcm^{-2}K{-1}, typical for this type of measurements, see Refs.[1,8,9]. Thus, Eq.[1] confirms the heating origin of the falling part of the R(P) dependence in Fig.1(a) and suggests that the IVC by Ref.[10] will be linear above and below T_c if the heating artefacts are removed.

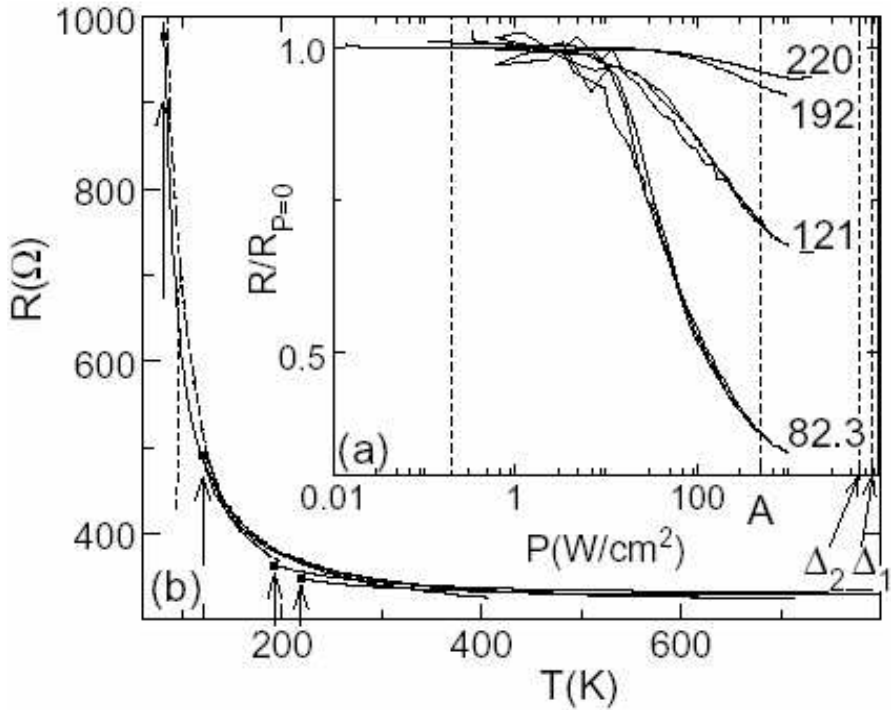

Fig.1 (a): Typical nonlinear IVC, measured at different T_B above and below T_c=93K, are re-plotted as a sample resistance, R=V/I, normalised by its value at P=0, versus the heat load, P=IV/A. T_B are shown in the figure at the corresponding curves and also by the arrows in Fig.1(b); the solid dots in Fig.1(b) represent R(P=0). The characteristic heat loads which build the IJT gaps and the point 'A' in the comment's IVC are shown by the broken lines and the axis labels.
Fig.1(b) Compares the measured R(T) shown by the broken line and the ones calculated with Eq.(1) from the nonlinear IVC using one and the same heat transfer coefficient  h=32Wcm^{-2}K{-1} for the data taken at T_B spanned over 140K.
.
As is seen from Fig.1a, the critical load drops with temperature radically, as does the range of loads where intrinsic features dominate. For this reason a study of intrinsic response becomes enormously complicated at helium temperatures where the extrinsic features dominate almost throughout the range of loads where the brush-like IVC exists. Indeed, the IJT spectra taken at T_B=4.3K often reveal rather nonlinear branches. Below I will consider such a case and, as an illustration, will show that heating is a likely cause of the nonlinear response in the both structures discussed in the comment. Henceforth these structures will be referred as 1 and 2 in accordance with the numbers of the corresponding figures.

As is seen from Fig.1a, where the heat loads related to the characteristic IVC points are indicated by the axis labels and broken lines, the point 'A' in IVC-2 is built by P('A')=500W/cm^2 which exceeds the critical load for the lowest $T_B$ in Fig.1(a) by more than two orders. Thus, the corresponding part of IVC-2 will be caused by heating and definitely belong to the falling part of R(P) even if this IVC-2 were taken at $T_B$=82K. As, however, the IVC-2 is taken at lower $T_B$=5.6K, P(A) should be compared with a $P_c$ value at least 20 times smaller (see Refs.[1,3] for details). Thus, the range of loads where extrinsic features prevail spans over 3.5-4 orders below P('A'), covering the entire IVC-2.

It is worth noting that the self-heating cause of IVC-2 is indirectly admitted by the author of the comment who declares that IVC-1 represents "the case of extreme self-heating". Indeed, the heat loads which build the IJT gaps in these samples are practically the same and so there are no valid reasons to expect the self-heating to be radically different, see Ref.[11].

Thus, the IVC-2 supports neither the claim that "the self-heating along the branches is negligible" nor that "the genuine interlayer IVC's are strongly nonlinear". The last major claim of the comment, that the branches in the brush "are perfectly periodic" is also at odds with the experiment because neither the genuine branches described by Eq(2), nor the nonlinear ones, advocated by the author of the comment, obey the definition of periodicity:

$$F(x+a)=F(x), a=const. \qquad Eq.(3)$$

Furthermore, the real branches are only approximately regular and frequently reveal extra branches with irrational numbers.

To conclude, neither the argumentation nor the conclusions of the comment by V. Krasnov are experimentally justified. Contrary to the comment's claims, Ref.[1] addresses the genuine IVC experimentally and shows that at sufficiently high heat loads the heating-induced IVC nonlinearities exceed the plausible intrinsic ones, eg. of Eq.(2), so radically that the latter might be safely ignored. As is seen from Fig.1b and numerous similar findings reported by Refs.[1,4,5,7,8,9], Newton's Law of Cooling (1701) and Ohm's law describe the experimental IVC quantitatively using the normal state resistance of the same sample only.

References

[1] V.N.Zavaritsky, Phys.Rev.B72,094503(2005).
[2] A.V.Gurevich and R.G. Mints, Rev.Mod.Phys 59, 941(1987).
[3] V.N. Zavaritsky, J.Vanacen, V.V.Moshchalkov, A.S.Alexandrov, cond-mat/0308256; Physica C 404, 444 (2004).
[4] V.N.Zavaritsky, M. Springford, A.S. Alexandrov, cond-mat/0006089; Europhys. Lett, 51, 334 (2000); Physica B 294-295, 363 (2001) ;V.N.Zavaritsky, Sov Zh. Exp and Theor. Physics 121, 1(2002); JETP 94, 933 (2002).
[5] The area independence of heating effects observed by Refs.[1,7] was strongly supported by the recent comment, which addressed the heating cause of IJT spectra and discovered that practically the same heat loads (P\sim10kW/cm^2) build the IJT gap in Bi2212 structures of vastly different area 1<A<30mcm^2. This comment entitled "Intrinsic tunnelling or wishful thinking?" was submitted to PRL on 22.08.05. The comment also shows that the principal theoretical failure of the authors of PRL94,077003 consists in their belief that heat escapes into the poorly conducting substrate *only,* hence ignoring the principal channel of heat escape (through the metal electrode of enhanced area and convection). The comment briefly addresses the other major experimental faults of the authors of this letter; in particular, it shows that the nonequilibrium thermometry by this group is unreliable as it ignores the heat load dependent thermal lag.
 [6] A. Yurgens, D. Winkler, T. Claeson, S. Ono, and Y.Ando, cond-mat/0309131, cond-mat/0309132; see also Phys. Rev. Lett. 92, 259702 (2004).
[7] V.N.Zavaritsky, J Supercond. 15, 567 (2002)
[8] V.N. Zavaritsky, Physica C404, 440 (2004)
[9] V.N. Zavaritsky, cond-mat/0306081; see also Phys. Rev.Lett. 92, 259701 (2004).
[10] V. M. Krasnov, A.Yurgens, D.Winkler, P.Delsing,T.Claeson, cond-mat/0002172 ; Krasnov, cond-mat/0601101 ; see also PRL84,2000,5860
[11] Detailed quantitative analysis of the actual overheating in these samples is not possible in the absence of reliable $R_N(T)$ or h,.